\renewcommand\k{\kappa}

\newcommand{\diracslash}[1]{#1\llap{/\kern2pt}}

\def\bearr{\begin{eqnarray}}
\def\eearr{\end{eqnarray}}

\newcommand{\bea}{\begin{eqnarray}}
\newcommand{\eea}{\end{eqnarray}}

\documentclass[preprint,secnumarabic,floats,amssymb,nobibnotes,nofootinbib,aps,prc,showpacs,tightenlines]{revtex4}
\usepackage{shrthnds}
\usepackage{graphicx}
\usepackage{amsmath}
\usepackage{latexsym}
\usepackage{epsfig}
\usepackage{subfigure}
\usepackage[dvipsnames]{xcolor}
\begin{document}

\title{Higgs Inflation in $f(\Phi,R)$ Theory}
\author{
Girish Kumar Chakravarty$^{}$\footnote{email: girish20@prl.res.in},
Subhendra Mohanty$^{}$\footnote{email: mohanty@prl.res.in} and
Naveen K. Singh$^{}$\footnote{email: naveenks@prl.res.in} }
\affiliation{ Theory Division, Physical Research Laboratory,
Navrangpura, Ahmedabad 380 009, India}
\date{\today}

\begin{abstract}
We generalize the scalar-curvature coupling model ${\xi \Phi^2 R}$ of Higgs inflation to $ {\xi \Phi^a R^b} $ to study inflation.
  We compute the amplitude and spectral index of curvature perturbations
  generated during inflation and fix the parameters of the model by comparing these with the Planck$+$WP data. We find that
 if the scalar self coupling $\lambda$ is in the range $(10^{-5}-0.1)$, parameter $a$ in the range $(2.3 -3.6)$ and $b$ in the range
 $(0.77-0.22)$ at the Planck scale, one  can have a viable inflation model even for $\xi \simeq 1$. The tensor to scalar ratio $r$ in this
 model is small and our model  with scalar-curvature couplings is not ruled out by observational limits on $r$ unlike the
 pure  $\frac{\lambda}{4} \Phi^4$ theory. By requiring the curvature coupling parameter to be of order unity, we have
 evaded the problem of unitarity violation in scalar-graviton scatterings which plague the $\xi \Phi^2 R$ Higgs inflation models.
 We conclude that the Higgs field may still be  a good candidate for being the inflaton in the early universe if one
 considers higher dimensional curvature coupling. \\
 
\hspace{-.5cm} Keywords: Higgs Inflation; CMB spectrum; Non-minimal coupling; Jordan frame; Einstein frame;
Perturbations.

\end{abstract}

\pacs{98.80.Cq, 14.80.Bn}

\maketitle

\section{Introduction}
The idea that the universe through a period of exponential expansion, called inflation \cite{guth,starobinsky1,starobinsky2,kazanas,sato,linde1,linde2,linde3,albrecht}
has proved useful for solving the horizon and flatness
problems of standard cosmology and in addition providing an explanation for the scale invariant super-horizon
perturbations which are responsible of generating the CMB anisotropies and formation of structures in the universe.
A successful theory of inflation requires a flat potential where a scalar field acquires a slow-roll over a sufficiently
long period to enable the universe to expand by at least 60 e-foldings during the period of inflation. There is a wide variety
of particle physics models which can provide the slow roll scalar field 'inflaton' for inflation \cite{Lyth:1998xn}.
From the observations of CMB anisotropy spectrum  by COBE and WMAP \cite{WMAP9} it is not yet possible to pin down a
specific particle physics model as the one responsible for inflation. In the light of recent discoveries by CMS \cite{cms} and
ATLAS \cite{Atlas} it is of interest to consider the Standard Model Higgs boson as the candidate for inflaton. On the face of it
the idea does not work as the inflaton quartic coupling should be of the order $\lambda \sim 10^{-12}$ to explain the amplitude of
CMB perturbations measured by WMAP \cite{WMAP9} while the 125 GeV Higgs has a quartic coupling $\lambda \sim 0.13$ at the electroweak
scale which can however go down to smaller values at the Planck
scale due to renormalization \cite{Sher,Froggatt,Espinosa,Schrempp,Holthausen,Degrassi}. However just from the standard model
renormalization one cannot have the Higgs coupling  $\lambda \sim 10^{-12}$ over the entire range of the rolling field (10-1 )$M_P$ during
inflation and the standard slow roll inflation with a
Higgs field does not give the observed amplitude and spectrum of density perturbations \cite{Isidori:2007vm}. If the Higgs and
top mass are fine tuned then there can be a small kink in the Higgs potential and the universe trapped in  this false vacuum can
undergo a period of inflation \cite{Masina:2011aa,Masina:2011un,Masina}. 

Later a way out of fine tuning the scalar self coupling to unnaturally small values was 
 found out \cite{Spokoiny:1984bd,Futamase:1987ua,Salopek:1988qh,Fakir:1990eg}
and it was shown that if one couples
the scalar field to the Ricci scalar $\xi \Phi^2 R$ then the effective potential in the Einstein frame becomes a slow 
roll one with the effective scalar coupling being $\lambda/\xi^{2}$ and the amplitude of the density perturbations constrain this ratio rather than $\lambda$, 
hence $\xi$ can be increased as large as required to get the desired self-coupling $\lambda$. Density perturbations from inflation in the curvature coupled theories were calculated 
in \cite{Kaiser:1994vs, Komatsu:1997hv}. The equivalence of the density perturbation in Jordan and Einstein frame  was shown by Komatsu and Futamase \cite{Komatsu:1999mt} who 
also calculated the tensor perturbations and showed that the tensor to scalar ratio is generically small in $\xi \Phi^2 R$ model.

Bezrukov and Shaposhnikov \cite{Bezrukov} revived the large curvature coupling model to motivate the idea that the standard model Higgs
 field could serve as the inflaton in the early universe. The
amplitude and spectral index of density perturbations observed by WMAP can be generated by the Higgs field with self coupling $\lambda\sim 0.1$ and curvature
coupling  $\xi\sim 10^4$ \cite{Bezrukov,Barvinsky:2008ia,Bezrukov1,Bezrukov2,DeSimone:2008ei,Barvinsky:2009fy}. This large value of $\xi$ needed however is seen as a problem as
 at the time of inflation  the Higgs field is at the Planck scale and hence graviton-scalar scatterings due to the curvature coupling of the scalar would become
non-unitary \cite{Hertzberg:2010dc}. Ways of solving the unitarity violation problem in the Higgs inflation models have been explored in
\cite{Barvinsky:2009ii,giudice,lerner1,bauer,Qiu}. 

In this paper we  assume that the dominant interaction between Higgs field and gravity is through 
operators of the form
\be
{\cal L}=\frac{\xi ({\cal H}^\dagger {\cal H})^{a/2} R^b}{M_p^{a+2b-4}}. \label{nonminimal}
\ee
This form (\ref{nonminimal}) of Higgs Curvature interaction has been mentioned in the Ref. \cite{Atkins:2010yg}.
The complete dynamics of the Higgs field involves the role of the Goldstone modes as has been studied in detail in
\cite{Mooij:2011fi,Kaiser:2010yu,Greenwood:2012aj}. The multifield dynamics of the Goldstone modes gives rise to sizable non-gaussianity.
We will study the dynamics of the Higgs mode and impose a charge conservation and CP symmetry such that the Goldstone modes of the
Higgs field do not acquire vevs. We will take the background Higgs field to be
\be
{\cal H} =\begin{pmatrix}
  0  \\
  \Phi
 \end{pmatrix}
\ee
where $\Phi$ is the Higgs mode with  mass 126 GeV.
Our inflation model falls in the class of inflation in $f(\Phi,R)$ theories studied in Ref. \cite{shinji}. Our motivation is that we use the
 Higgs quartic coupling $\lambda ({\cal H}^\dagger {\cal H})^2$ where the standard model value of $\lambda(\mu\sim M_P)$ can
 lie in the range $\lambda = (10^{-5}-0.1)$ depending on the value of  top quark mass  \cite{Holthausen,Degrassi} or
 on new physics \cite{Chakrabortty:2012np}. We take curvature coupling
 $\xi$ to be unity and check the possibility of generating the observed density perturbations from Higgs inflation by varying parameters
 $a$, $b$ and $\lambda$. The non minimal coupling $\xi$ has been taken unity in order to improve the unitarity behaviour which increases the 
natural cutoff scale $\Lambda$ from $\Lambda \simeq \frac{M_{p}}{\xi} \simeq 10^{15}$ to $\Lambda \simeq M_{p} \simeq 10^{19}$.

We derive the curvature perturbation during inflation in two different ways. We derive the perturbations of modified Einsteins field equation in the
Jordan frame in presence of the  Higgs-curvature interaction terms and derive the amplitude and spectral index of
curvature perturbation. We find that to generate the Planck$+$WP  preferred amplitude $\Delta_{\mathcal R}^2= 2.1955^{+0.533}_{-0.585}\times 10^{-9} $
and spectral index $n_s= .9603 \pm .0073$  \cite{Planck_group} for $\lambda= 10^{-3}$ we should 
have $ a \sim 3.02, b \sim 0.49$ ( and for $\lambda=0.1$ we need $a \sim 3.56 , b \sim .22$). In these fits we take
$\xi = 1$. 

In the $\xi \Phi^2 R$ theory we can always make a conformal transformation to the Einstein frame so one can compute the density perturbations either in 
Einstein frame or Jordan frame and the gauge invariant curvature perturbations should be same in both the frames \cite{Kaiser:1994vs}.
In our case with the $\xi \Phi^a R^b$ coupling we find that no conformal transformation exists which can in general remove this term (i.e go to an Einstein frame).
We find that in the general $\xi \Phi^a R^b$ theory such a conformal transformation is only possible if the metric is quasi-de Sitter.
 The accurate comparision with the experimental data should be made however with the Jordan frame results.

Calculation of the curvature perturbation 
in both Einstein and Jordan frame for the $\xi \Phi^2 R$ theory has been done previously 
in Ref. \cite{Kaiser:1994vs,Gong:2011qe,prokopec,Basak:2012pc}. In  section (\ref{s_jordan}) we derive the curvature perturbations and
tensor perturbation in our theory in the Jordan frame and in  section (\ref{s_einstein}) we make a conformal
transformation to go to the Einstein frame and compute the curvature perturbations.   Finally in the last section (\ref{concl}) we compare the results of the two
 frames and discuss the viability of our considered Higgs inflation model.

\section{Calculation In The Jordan Frame}\label{s_jordan}

In this section we introduce a scalar-gravity interaction term  $f(\Phi,R)$ in the  action and
calculate physical quantities related to the inflationary density perturbations such as the spectral index, curvature
perturbation  and tensor-to-scalar ratio. We start with the action for a scalar field interacting with gravity of the form
\ba
S=\int d^4x\sqrt{-g}\Bigg[-\frac{f(\Phi,R)}{2\kappa^2} + \frac{1}{2}g^{\mu\nu}\partial_{\mu}\Phi\partial_{\nu}\Phi
+V(\Phi)\Bigg] \label{action_J} \ ,
\ea
where we take,
\ba
\frac{1}{\kappa^2}f(\Phi,R)= \frac{1}{\kappa^2}R + \frac{\xi \Phi^a R^b}{M_p^{a+2b-4}} \ \ ; \ \ V(\Phi) = \frac{\lambda \Phi^4}{4} \label{eqn_f} \ ,
\ea
 where $\kappa^2 =1/M_p^2$ and $\xi$ is a dimensionless coupling constant. Varying the action (\ref{action_J}) w.r.t $g^{\mu\nu}$ and $\Phi$ we obtain
the field equations,
\ba
G_{\mu\nu}=F R_{\mu\nu}-\frac{1}{2}fg_{\mu\nu}- \triangledown_{\mu}\triangledown_{\nu}F+g_{\mu\nu}\square F&=&
\kappa^2 \left(\triangledown_{\mu}\Phi \triangledown_{\nu}\Phi - \frac{1}{2}
g_{\mu\nu} \triangledown^{\rho}\Phi \triangledown_{\rho}\Phi - V g_{\mu\nu}\right) \label{eqn_J1}, \\
\square{\Phi}&=& V_{,\Phi} - \frac{f_{,\Phi}}{2\kappa^2},\label{eqn_J2} \
\ea
where $F=\partial{f}/\partial{R}=1+\frac{\xi b \Phi^a R^{b-1}}{M_p^{a+2b-2}}$.
\subsection{ Background quasi de-Sitter solution}

 For the unperturbed background FRW metric $diag(-1, a^2(t), a^2(t), a^2(t))$ and scalar field $\Phi=\phi(t)$, the above Eqs. (\ref{eqn_J1}) and
(\ref{eqn_J2}) reduce to the form
\ba
3 F H^2 + \frac{1}{2}\left(f-RF\right)+ 3H\dot{F}&=&\kappa^2\left(\frac{1}{2}\dot{\phi}^2 + V(\phi)\right) \label{leading_J1}\\
- 2 F\dot{H} -\ddot{F} + H \dot{F}&=& \kappa^2 \dot{\phi}^2 \label{leading_J2}\\
\ddot{\phi} + 3H\dot{\phi} + V_{,\phi} - \frac{f_{,\phi}}{2\kappa^2} &=&0 \label{leading_J3}\ .
\ea
Now we assume the second term of $F$ i.e. $\frac{\xi b \phi^a R^{b-1}}{M_p^{a+2b-2}}$ is dominant for some values of
$a$ and $b$. We find this assumption to be valid while solving numerically for the values of $a$ and $b$ in our model which give rise
to the experimentally observed density perturbations as discussed in the section (\ref{data}).
 From Eq. (\ref{leading_J1}), under this assumption and considering the slow roll parameters which are
 defined in Eq. (\ref{slow_rolls}) as small, the Hubble parameter  in the Jordan frame turns out to be of the form
\ba
H = \frac{\lambda^{\frac{1}{2b}}}{\sqrt{12}\big[\xi(2-b)\big]^{\frac{1}{2b}}}
\left(\frac{\phi}{M_p}\right)^{\frac{4-a}{2b}} M_p \label{Hubble_J} \ .
\ea
From Eq. (\ref{leading_J3}) under the slow roll assumption we get
\ba
\dot \phi=-\frac{\lambda \phi^3}{3 H}\Big[1- \frac{a}{2(2-b)}\Big]. \label{phidot}
\ea

\subsection{Scalar field and metric perturbations}\label{scal_pert}

 Now we perturb Eqs. (\ref{eqn_J1}) and (\ref{eqn_J2}) by perturbing the scalar field $\Phi=\phi(t)+\delta \phi(x,t)$ and the metric as
\ba
ds^2= -(1+2\alpha)dt^2 - 2  a(t) (\partial_i \beta)dt dx^i +  a^2(t)\left(\delta_{ij}(1 + 2\psi) + 2
\partial_i\partial_j \gamma \right) dx^i dx^j ,
\ea
where, $\alpha$, $\psi$, $\beta$ and $\gamma$ are scalar perturbations. We derive the Einstein equations for
the $f(R, \phi)$ theory \cite{hwang,hwang1} keeping the first order terms in the metric and scalar field perturbations.
 The component $\delta G_{00}$ is given by

\ba
\frac{\triangle}{ a^2(t)}\psi + H A = \frac{-1}{2F} \Bigg[ \left(3H^2+3\dot{H} +\frac{\triangle}{ a^2(t)}\right)\delta F
- 3 H\dot{\delta F} + \frac{1}{2} \left( 2 \kappa^2 V_{, \phi} -f_{, \phi}\right)\delta \phi \nonumber \\
+ \kappa^2 \dot{\phi}\dot{\delta \phi} +(3 H \dot{F} -\kappa^2 \dot{\phi}^2)\alpha + \dot{F}A\Bigg] , \label{pert1}
\ea
and taking the difference $\delta G^i_i-\delta G_0^0$
we get
\ba
\dot{A} + 2H A + \left(3 \dot{H} + \frac{\triangle}{ a^2(t)}\right) \alpha = \frac{1}{2 F} \Bigg[
3 \ddot{\delta F} + 3 H \dot{\delta F} - \left(6 H^2 + \frac{\triangle}{a^2(t)}\right)\delta F
+ 4 \kappa^2 \dot{\phi}\dot{\delta \phi} \nonumber \\
+\left(-2 \kappa^2 V_{,\phi} + f_{,\phi}\right)\delta \phi - 3 \dot{F}\dot{\alpha}-\dot{F}A\nonumber \\
-\left( 4 \kappa^2 \dot{\phi}^2 + 3 H \dot{F} + 6 \ddot{F}\right)\alpha \Bigg] \label{pert4}
\ea
where $A=3(H\alpha -\dot{\psi}) -\triangle \chi /a^2(t)$ and $\chi =  a(t) (\beta + a \dot{\gamma})$. Here, in
arriving the Eqs. (\ref{pert1}) and (\ref{pert4}), the leading order Eqs. (\ref{leading_J1}) and (\ref{leading_J2}) are also used.
The other components  $\delta G_{i0}$ and $\delta G_{ij}$ ($i\neq j$) of the first order perturbed  Einstein equation  can be written as
\ba
H \alpha - \dot{\psi} &=& \frac{1}{2 F} \Bigg [ \kappa^2 \dot{\phi} \delta {\phi} + \dot{\delta F} - H \delta F
-\dot{F}\alpha \Bigg] , \label{pert2}
\ea
and
\ba
\dot {\chi} + H \chi -\alpha -\psi &=& \frac{1}{F} \left( \delta F - \dot{F} \chi \right)  \label{pert3}
\ea
respectively. The equation of motion of scalar perturbation is
\be
\ddot{\delta \phi} +  3H\dot{\delta \phi} + \Bigg[-\frac{\triangle}{ a^2(t)} + \left( \frac {2 V_{,\phi}-
f_{,\phi}/{\kappa^2}}{2}\right)\Bigg]\delta \phi = \dot{\phi}\dot{\alpha} + \left(2 \ddot{\phi}
+ 3 H \dot \phi\right)\alpha + \dot{\phi}A +\frac{1}{2} F_{,\phi}\left(\frac{\delta R}{\kappa^2}\right) .  \label{eom_pert}
\ee
where
\be
\delta R = -2 \Bigg[\dot{A}+4AH +\left(\frac{\triangle}{a^2(t)}+3 \dot{H}\right)\alpha +2 \frac{\triangle}{ a^2(t)}\psi \Bigg]
\ee
Now we analyze the curvature perturbation $\mathcal{R}=\psi-H\delta \phi/\dot{\phi}$ by choosing a gauge where $\delta \phi=0$ and $\delta R=0$. This
 sets $\mathcal{R} =\psi$ and moreover we
have $\delta F=0$ via $\delta F = \left(\partial F/\partial \phi\right) \delta \phi +
\left(\partial F/\partial R\right) \delta R$. Under this gauge the Eq. (\ref{pert2}) gives,
\ba
\alpha = \frac{\dot{\mathcal{R}}}{H + \dot{F}/(2F)} \label{eqR1}
\ea
and hence from Eq. (\ref{pert1}) we get
\ba
A = - \frac{1}{H + \dot{F}/(2F)} \left(\frac{\triangle}{ a^2(t)}\mathcal{R}
+ \frac{\left(3 H \dot{F}-\kappa^2\dot{\phi}^2\right)\mathcal{\dot{R}}}{2F\left(H + \dot{F}/(2F)\right)} \right) \label{eqR2} .
\ea
Using  Eq. (\ref{leading_J2}) and Eq. (\ref{pert4}), we obtain
\ba
\dot{A} + \left(2H + \frac{\dot{F}}{2 F}\right) A + \frac{3 \dot{F}}{2 F}\dot{\alpha}
+ \left(\frac{3  \ddot{F} + 6 H \dot{F} + \kappa^2 \dot{\phi}^2}{2 F} + \frac{\triangle}{ a^2(t)}\right)\alpha =0 . \label{eqR3}
\ea
 Now we may write the differential equation for
curvature perturbation by using the above Eqs. (\ref{eqR1}), (\ref{eqR2}) and (\ref{eqR3}) as
\ba
\ddot{\mathcal{R}} + \frac{(a^3(t) Q_s)\dot{}}{ a^3(t) Q_s} \dot{\mathcal{R}} + \frac{k^2}{a^2(t)}\mathcal{R}=0, \label{eqR}
\ea
where,
\ba
Q_s = \frac{\dot{\phi}^2 + 3 \dot{F}^2/(2\kappa^2 F)}{\left(H+\dot{F}/(2F)\right)^2} \label{eqn_Qs}.
\ea
In arriving Eq. (\ref{eqR}),  Eq. (\ref{leading_J2}) is again used. Now one may re-write the Eq. (\ref{eqR}) in terms of
variables $\omega = a \sqrt{Q_s}$ and $\sigma_k = \omega \mathcal{R}$ as
\ba
\sigma_k'' + \left(k^2 - \frac{\omega''}{\omega}\right)\sigma_k = 0 , \label{eqR_final}
\ea
where prime denotes the derivative with respect to the conformal time defined as
$d\eta = dt/a(t)$ and
\ba
\frac{\omega''}{\omega} = \frac{a''(t)}{a(t)} + \frac{a'(t)}{a(t)} \frac{Q_s'}{Q_s} + \frac{1}{2}\frac{Q_s''}{Q_s} -\frac{1}{4} \left(\frac{Q_s'}{Q_s}\right)^2
\ea
under quasi de-Sitter expansion $a(\eta)=\frac{-1}{H \eta (1-\epsilon_1)}$ and hence $\frac{a''(t)}{a(t)}=\frac{1}{\eta^2}\big[2+3\epsilon_1 \big]$ and $ a'(t)/a(t)=a(t) H $.
Therefore we have
\ba
\frac{\omega''}{\omega} = \frac{1}{\eta^2} \Big[\nu_{\mathcal{R}}^2 - \frac{1}{4}\Big]
\ea
where
\ba
\nu_{\mathcal{R}}^2 = \frac{9}{4}\Big[1+\frac{4}{3}\left(2\epsilon_{1} + \epsilon_{2} - \epsilon_{3} + \epsilon_{4}\right)\Big].
\ea
In arriving at the above expression we have defined
\ba
\epsilon_1=-\frac{\dot{H}}{H^2} \ , \ \epsilon_2 =\frac{\ddot{\phi}}{H \dot\phi} \ , \ \epsilon_3=
\frac{\dot F}{2 H F} \ , \ \epsilon_4=\frac{\dot E}{2 H E} \ \label{slow_rolls};
\ea

\ba
\ E = F+ \frac{3\dot{F}^2}{2
\kappa^2\dot{\phi}^2} =\frac{Q_s (1+\epsilon_3)^2 }{\dot{\phi}^2/(F H^2)} \ .
\ea
Here $\epsilon_i$ are slow roll parameters and $\dot{\epsilon_{i}}$ terms have been neglected.
Equation (\ref{eqR_final}) then has solutions in the Hankel functions of order $\nu_R$
\be
\sigma=\frac{\sqrt{\pi |\eta|}}{2} e^{i (1+2 \nu_{\mathcal R})\pi/4} \left[ c_1\, H_{\nu_{\mathcal R}}^{(1)}(k |\eta|)+
c_2\, H_{\nu_{\mathcal R}}^{(2)}(k |\eta|) \right]
\ee
 Applying the Bunch-Davies boundary condition $\sigma(k \eta \rightarrow -\infty)=e^{i k \eta}/\sqrt{2 k}$  we fix the
 integration constants $c_1=1$ and $c_2=0$. Using the relation $H_\nu(k |\eta|)= \frac{-i}{\pi} \Gamma(\nu) \left(\frac{k |\eta|}{2} \right)^{-\nu}$ for
 the super-horizon modes $k \eta \rightarrow 0$, we obtain the expression for the power spectrum for curvature perturbations is defined as
\be
{\mathcal P}_{\mathcal{ R}}=\frac{ 4 \pi k^3}{(2 \pi)^3} |{\mathcal R}|^2 \equiv {\Delta }_{\mathcal R}^2 \left(\frac{k}{a(t)H} \right)^{n_{\mathcal R}-1}
\ee
The amplitude of the curvature power spectrum turns out to be
\be
{\Delta}_{\mathcal{R}}=\frac{1}{\sqrt{Q_s}}\left(\frac{H}{2\pi}\right)
\ee
and the spectral index is
\be
n_{\mathcal R}-1=3-2\nu_{\mathcal{R}}\simeq -4\epsilon_1-2\epsilon_2+2\epsilon_3-2\epsilon_4 \simeq - 6 \epsilon_1 \label{spectral} \ .
\ee

Using slow roll parameters, Eq. (\ref{eqn_Qs}) can be simplified to the form $Q_s \simeq 6 F \epsilon_3^2 M_p^2$ with $\frac{\kappa^2 \dot{\phi}^2}{
F H^2}<< 6 \epsilon_3^2 $ which will be justified later in  section (\ref{data}). In our model of $f(\Phi, R)$ coupling
we find  $\epsilon_1\approx -\epsilon_3$, $\epsilon_2\approx-\epsilon_4$ and these
relations are used in the calculation of perturbation amplitude and spectral index. Plugging the values $H$ and $\dot{\phi}$ from Eqs.
(\ref{Hubble_J}) and (\ref{phidot}) into Eq. (\ref{slow_rolls}), the slow roll parameters can be written as
\ba
\epsilon_1 &=&  b^{-1} (a-4) (2-b)^{(1-b)/b} (a+2 b-4)\lambda ^{(b-1)/b}
\xi ^{1/b} \left(\frac{\phi}{M_p}\right)^{\frac{a+2 b-4}{b}} \label{sl1} \\
\epsilon_2 &=& b^{-1}(a+6b-4) \left(2-b\right)^{(1-b)/b}(a+2b-4) \lambda^{(b-1)/b}\xi^{1/b}\left(\frac{\phi}{M_p}\right)^{\frac{a+2 b-4}{b}} \label{sl2}  .
\ea
For our model, we can write the expressions for the amplitude of power spectrum  and the number
of e-folding as
\ba
\Delta_{\mathcal{R}}^2 = \frac{b [(2-b)/\lambda]^{3-\frac{4}{b}}  M_{p}^{8+\frac{4 (a-4)}{b}}
 \xi ^{-\frac{4}{b}} \phi^{-\frac{4 (a+2 b-4)}{b}}}{288 (a-4)^2 (a+2 b-4)^2 \pi ^2}
\label{DeltaJ}
\ea
and
\ba
N_J= \int_{\phi_J}^{\phi_f} \frac{H}{\dot \phi} d \phi =  \frac{b [(2-b)/\lambda]^{\frac{b-1}{b}} \xi^{-\frac{1}{b}}}{2(a+2b-4)^2}\left(\frac{\phi}{M_p}\right)^{\frac{4-a-2b}{b}}
\Bigg|^{\phi_J}_{\phi_f} \label{Nj}
\ea
respectively.  Here  $\phi_J$ and $\phi_f$ are the values of scalar field $\phi$
at the beginning and the end of inflation respectively.

\subsection{Tensor Perturbations}
We define the perturbation of metric as follows
\ba
g_{\mu\nu}= \bar{g}_{\mu\nu} + h_{\mu\nu} \ \ \mbox{and} \ \ g^{\mu\nu}= \bar{g}^{\mu\nu} + h^{\mu\nu}, \label{metric_pert1}
\ea
where $\bar{g}_{\mu\nu}$ is background metric and
\ba
h^{ij}=-\frac{1}{a^4(t)} h_{ij}, \ h^{i0}=\frac{1}{a^2(t)} h_{i0} , \  h^{00}= -h_{00}. \label{metric_pert2}
\ea
Now to get the equation of tensor pertubation,  we set $h_{i0}=h_{00}=0$ in the calculation.
From the decomposition theorem, the non zero spatial components $h_{ij}$ are traceless and divergenceless, i.e., 
\ba
h_{ii}=0, \ \ \ \partial_{i}h_{ij}=0. \label{decomp}
\ea
Using Eqs. (\ref{metric_pert2}) and (\ref{decomp}), we obtain 
\ba
\delta R_{00}&=&0, \delta R_{i0}=0,
\ea
\ba
\delta R_{ij}&=&-\frac{1}{2 a^2(t)}\triangledown^2 h_{ij} +\frac{1}{2}\ddot{h}_{ij} -\frac{\dot{a}}{2a}\dot{h}_{ij} 
+ 2 \left(\frac{\dot a}{a}\right)^2 h_{ij}, \ \ \delta R =0.
\ea
So, perturbing Eq. (\ref{eqn_J1}), we obtain
\ba
\frac{1}{2}F a^2\ddot{D}_{ij} + \left(\frac{1}{2}\dot{F}a^2 + \frac{3}{2}a\dot{a}F\right)\dot{D}_{ij} -\frac{F}{2}\triangledown^2 D_{ij}
=\Big[2 \frac{\dot{a}}{a} \dot{F} -2F \left(\frac{\dot a}{a}\right)^2 - \frac{\ddot a}{a} F + \frac{f}{2} \nonumber \\ 
+ \ddot{F} + \frac{\kappa^2}{2}\left(\dot \phi^2 - 2 V\right)\Big]a^2 D_{ij}, \label{eqn_t}
\ea
 where $D_{ij}= h_{ij}/a^2$. The right hand side of Eq. (\ref{eqn_t}) vanishes by Eqs. (\ref{leading_J1}) and (\ref{leading_J2}). Thus
 we have
 \ba
 \ddot{D}_{ij} + \frac{(a^3 F)\ \dot{}}{a^3 F}\dot{D}_{ij} + \frac{\kappa^2}{a^2}D_{ij}=0. \label{tensor_eqn}
 \ea
 In the terms of polarization tensors $e_{ij}^{1}$ and $e_{ij}^2$, the tensor $D_{ij}$ is written as
 \ba
 D_{ij}= D_1 e_{ij}^1 + D_2 e_{ij}^2.
 \ea
For gravity wave propagating in $\hat z$ direction, the components of polarization tensor are given by
\ba
e_{xx}^1=-e_{yy}^1=1, \ \ e_{xy}^2=e_{yx}^2=1, \ \  e_{iz}^{1,2}=e_{zi}^{1,2}=0.
\ea
So the Eq. (\ref{tensor_eqn}) can be written as 
\ba
\ddot{D}_{\lambda} + \frac{(a^3 F)\ \dot{}}{a^3 F}\dot{D}_{\lambda} + \frac{\kappa^2}{a^2}D_{\lambda}=0, \label{tensor_eqn1}
\ea
where $\lambda \equiv1,2$. Now substituting $z=a \sqrt{F}$ and $v_k=z D_{\lambda} M_P/\sqrt{2}$, we get
\ba
v_{\lambda}''+ \left(k^2-\frac{z''}{z}\right)v_{\lambda}=0, \label{tensor_eqn2}
\ea
where, prime $^{\prime}$ is derivative with respect to conformal time.
Summing over all polarization states, the Eq. (\ref{tensor_eqn2}) provides us the 
amplitude of power spectrum  of $D_{\lambda}$ as
 \ba
 P_T=4 \times \left(\frac{2}{M_p^2}\right) \frac{\k^3}{2\pi^2}\frac{1}{a^2 F} v_{\lambda}^2\simeq\frac{2}{\pi^2} \left(\frac{H}{M_P}\right)^2\frac{1}{F}.
 \ea
 So, the ratio of the amplitude of tensor perturbations to scalar perturbations $r$ for $f(\Phi,R)$ theories is given
  by  
\be
r\simeq\frac{8\kappa^2 Q_s}{F} \simeq  48 \epsilon_3^2  \ .
\ee

\section{Calculation in the Einstein Frame} \label{s_einstein}
Starting with the considered action
\ba
S &=& \int d^4 x \sqrt{-g}\Big[-\frac{{M_p}^2}{2} R\left(1 + \frac{\xi \Phi^a R^{b-1}}{M_p^{a+2b-2}}\right) \nonumber \\
&+& \frac{1}{2}\partial_\mu \Phi \partial^\mu \Phi + \frac{\lambda \Phi^4}{4}\Big] \label{action1}
\ea
 we perform a conformal transformation of
the metric $g_{\mu\nu}$ to the Einstein frame metric $\tilde g_{\mu\nu}$which is defined as
\ba
\label{conf_trafo}
\tilde{g}_{\mu\nu}(x) &=& \Omega^2(x) g_{\mu\nu}(x)~,
\ea
where
\ba
\Omega^2  = 1 + \frac{\xi \Phi^a R^{b-1}}{{M_p}^{a+2b-2}} \label{Omega_eqn}\ .
\ea
The Ricci scalar transform as
\ba
\label{ricciscalar}
R &=& \Omega^2 \Big[ \tilde{R} + 6
 \frac{\tilde {\square} {\Omega} }{\Omega} - 12
 \frac{\tilde \partial^{\mu}\Omega \tilde \partial_{\mu}\Omega}{\Omega^2}\Big]~  . \label{ricci_trans}
\ea
For quasi de-Sitter space we can ignore the second  and third terms in the bracket in Eq. (\ref{ricci_trans}) which is justified in
the Eq. (\ref{check}). For this slow
roll case, we can write Eq. (\ref{Omega_eqn}) in Einstein frame as
\ba
\Omega^2= 1 + \frac{\xi^{1+\beta} \Phi^{\alpha}{\tilde{R}}^{\beta}}{M_p^{\alpha+2\beta}} \label{Omega_eqn1} \ ,
\ea
where, $\alpha= a/(2-b)$ and $\beta =(b-1)/(2-b)$.  Now we write the action (\ref{action1}) in term of new field $\phi_E$, which
is related to the field $\Phi$ by the relation
\ba
\frac{d \phi_E}{d\Phi}
= \frac{1}{\Omega^2}\left(\Omega^2 +  \frac {3 \alpha^2 \xi'^{2}}{2} \left(\frac{\Phi}{M_p}\right)^{2 \alpha -2}\right)^{1/2} ,\label{diff}
\ea
where $\xi'=\xi^{1+\beta}(\tilde{R}/M_p^2)^{\beta}$. This leads the action
in term of $\phi_E$ as follows
\ba
S_E  = \int d^4 x \left(- \frac{M_p^2}{2} \tilde R + \frac{1}{2} \tilde D^{\mu}{\phi_E}\tilde D_{\mu}\phi_E + U(\phi_E)\right), \label{Einstein}
\ea
where
\ba
U(\phi_E) =\frac{1}{\Omega^4} \frac{\lambda}{4}\Phi(\phi_E)^4 .\label{Pot1}
\ea
For $\Phi>> M_P/\xi'^{1/\alpha}$,  Eq. (\ref{diff}) can be integrated to give
\ba
\Phi= \frac{M_p}{\xi'^{1/\alpha}} \exp\left(\sqrt{\frac{2}{3}}\frac{\phi_E}{M_p \alpha}-\frac{1}{2}\right) \label{eqnh}.
\ea
Considering $\tilde g_{\mu\nu} = diag(-M^2(t),\tilde a^2(t),\tilde a^2(t),\tilde a^2(t))$ and varying the action (\ref{Einstein}) with
respect to $M(t)$ or $a(t)$ and setting $M=1$  in the final equation which corresponds FRW metric, we get the Friedmann equation
\ba
12 \tilde H^2  - \zeta^{-1} M_p^2\lambda \left(1+\frac{2\beta}{\alpha}\right) =0 \label{eq1} \ ,
\ea

 where
\ba
\zeta = 12^{4\beta/\alpha} \left(\frac{\tilde H^2}{M_p^2}\right)^{4\beta/\alpha} \xi^{\frac{4(1+\beta)}{\alpha}}
\exp\left(2 \sqrt{\frac{2}{3}}\frac{(\alpha-2)\phi_E}{\alpha M_p}\right).
\ea
Here we have neglected all the derivative terms of  Hubble parameter $\tilde H$. This corresponds to slow
roll condition, i.e., $\dot{\phi_E}^2$ is much smaller than potential term. We may write the Hubble parameter from
Eq. (\ref{eq1})  as
\ba
\tilde{ H}=M_p \frac{\left[\left(1+2 \beta/\alpha\right) \lambda \right]^{\frac{\alpha}{2(\alpha + 4\beta)}}}
{\sqrt{12} \ \xi^{\frac{2(1+\beta)}{\alpha+ 4 \beta}}}
\exp\Bigg[\sqrt{\frac{2}{3}}\left(\frac{2-\alpha}{\alpha+4\beta}\right) \frac{\phi_E}{M_p}\Bigg] \label{Hubble}\ .
\ea
 Now using Eq. (\ref{Hubble}) and (\ref{Pot1}) we obtain
\ba
U(\phi_E)=\frac{1}{4} M_p^4 \lambda^{\frac{\alpha}{\alpha+4\beta}}\xi^{-\frac{4(1+\beta)}{\alpha+4\beta}}
\left(1+\frac{2\beta}{\alpha}\right)^{- \frac{4\beta}{\alpha+4\beta}} \exp\Big[2 \sqrt{\frac{2}{3}}
\left(\frac{2-\alpha}{\alpha+4\beta}\right) \frac{\phi_E}{M_p}\Big] \label{Pot2}\ .
\ea
Here we have taken the approximation $\exp(\sqrt{\frac{2}{3}} \frac{\phi_E}{M_p}) >>1$ for $\phi_E>>M_p$.
 We now compute the spectral index and curvature perturbation using above potential (\ref{Pot2}).
The slow roll parameters for large $\phi_E>>M_p$ comes out to be
\ba
\epsilon= \frac{M_p^2}{2}\left(\frac{U'}{U}\right)^2=\frac{4}{3}\left(\frac{a+2b-4}{a+4b-4}\right)^2 \ ; \ \eta =
M_p^2 \left(\frac{U''}{U}\right)=\frac{8}{3}\left(\frac{a+2b-4}{a+4b-4}\right)^2
\ea
and the curvature perturbation
\ba
\Delta_{\mathcal R} &=& \frac{3 H^3}{2 \pi U'(\phi_E)} \nonumber \\
&=&\frac{1}{8 \sqrt{2} \ \pi }\left(\frac{y+2}{2y-x+4}\right)^{\frac{x+2y+4}{2 x}}
 \lambda^{\frac{2y-x+4}{2x}}\xi^{-\frac{2}{x}} \left(\frac{x}{y}\right)\exp\left(-\sqrt{\frac{2}{3}} \frac{y \phi_E}{x M_p} \right) \ , \label{curvp_E}
\ea
where $x= a+ 4b-4$ and $y=a+2b-4$.\\
The spectral index in the term of slow roll parameters is $n_s=1-6\epsilon + 2\eta$.\\
The number of e-folding is calculated as
\ba
N_{E} &=&\int^{{\phi_E}_0}_{{\phi_E}_e} \frac{U(\phi_E)}{U'(\phi_E)} d\phi_E \nonumber \\
&=& -\frac{1}{2}\sqrt{\frac{3}{2}}\left(\frac{x}{y}\right)\left(\frac{{\phi_E}_0 -{\phi_E}_e}{M_p}\right)
\label{N_E}
\ea
For  ${\phi_E}_0\sim 13 M_p$ and ${\phi_E}_e\sim 1 M_p$, the number
of e-folding is found to be around $60$. The slow roll parameters $\epsilon$, $\eta$ and the Hubble parameter $H$ are nearly independent of $\lambda$ and
are $\sim 0.02$, $\sim 0.04 $ and $ \sim 5.8 \times 10^{-5} M_p$ respectively. \\

Now from Eqs. (\ref{Omega_eqn}) and (\ref{eqnh}), we
can calculate the order of terms like $\ddot{\Omega}/\Omega$ and $(\dot\Omega/\Omega)^2$
for $\phi>>\frac{M_p}{\xi^{1/\alpha}}$. For $\lambda=10^{-3}$ and $\xi=1$,
\ba
\frac{\ddot{\Omega}}{\Omega} &\sim& \frac{U}{9 M_p^2} (\epsilon +\sqrt{3\epsilon} (\eta-\epsilon)) = 4.1 \times 10^{-11} M_p^2
\ \ \mbox{and}\nonumber \\
\left(\frac{\dot{\Omega}}{\Omega}\right)^2  &\sim& \frac{U}{9 M_p^2} \epsilon = 3.3 \times 10^{-11} M_p^2 \label{check}
\ea
respectively, whereas the value of curvature scalar $\tilde R =12 \tilde{H}^2$ at the same values of parameter is
$4.1 \times 10^{-8} M_p^2$. Thus our approximation (i.e. for quasi de-Sitter space we can ignore the second and third terms in the bracket in Eq. (\ref{ricci_trans})) made
is consistent and may be checked for other values of $a$ and $b$.

\begin{table}
\center
    \begin{tabular}{|l|l|l|l|l|l|}
        \hline
        $ \ \lambda$ & $ \ 0.1  $  &  $ \ 10^{-2}  $  &   $ \ 10^{-3} $ & $ \ 10^{-4} $ & $\ 10^{-5}$ \\ \hline
       \ a &$ \ 3.385  $ & $\ 3.026$  & $ \  2.735$ & $\ 2.494$&$\ 2.292$\\ \hline
       \ b & $ \ 0.277$ & $\ 0.439$  & $\ 0.571 $& $\ 0.679 $&$\ 0.770$ \\ \hline
       \ a+2b & $ \ 3.939 $ & $ \ 3.904$ & $ \ 3.877 $ & $ \ 3.852 $ & $ \ 3.832 $ \\ \hline 
        

    \end{tabular}
    \caption{The values of parameters (a,b) in the Einstein frame at $\phi_E=13 M_p$ with $\xi=1$ for different values of $\lambda$.}
    \label{values1_E}
\end{table}
We now use the measured values of these CMB anisotropy parameters to get the 
numerical values for the parameters  $(a,b,\xi,\lambda)$.

\section{Comparison with data }\label{data}
From the  Planck$+$WP measurements  \cite{Planck_group} we know that the curvature 
perturbation $\Delta_{\mathcal R}^2 = 2.195^{+0.533}_{-0.585} \times 10^{-9}$,
spectral index $n_{\mathcal R}= 0.9603 \pm 0.0073$ and the tesnsor to scalar ratio $r < 0.11 (95\% CL)$. 
For inflation to solve the horizon and flatness problems of standard hot big bang cosmology the number of e-foldings in the  Eintein
frame $N_E$ is required to be about 60.  From eqn. (\ref{N_E}) we see that to get $60$ e-foldings, the scalar field $\phi_E$ should 
roll from $13 M_p$ to $1 M_p$. We compute the curvature perturbation (\ref{curvp_E})  and spectral index in the Einstein frame and equate the expressions
with the Planck+WMAP values to compute the parameters $a$ and $b$ for different values of $\lambda$ and assume that the curvature coupling 
parameter $\xi=1$. Our results for the correlated set of parameters $\lambda, a,b$ at $\phi_E=13 M_P$ which give the measured values of
$\Delta_{\mathcal R}^2$ and $n_s$  are shown in the Table (\ref{values1_E}). We see that compared to the
$\xi \phi^2 R$ models with large $\xi$ the small deviations of $a$ and $b$ from 2 and 1 respectively can result in a large change  in $\xi$ which is 1 in our model
 compared to the earlier curvature coupling models where $\xi \sim 10^4$.

Next we equate the curvature perturbations  and spectral index in the Jordan frame from Eq. (\ref{DeltaJ}) and Eq. (\ref{spectral}) with the
Planck+WMAP data  to evaluate
the values of the parameters $\lambda, a$ and $b$ (keeping $\xi=1$) . The scalar field values $\Phi$ in the Jordan frame 
 corresponding to $\phi_E=1 M_p$ and $13 M_P$ for 
different values of $\lambda$ are displayed in Table (\ref{values3_J}). Using these values of the  range of the roll in 
$\Phi$ we see that the number of e-foldings $N_J$ in the Jordan frame, corresponding to $N_E=60$ is $N_J\sim 830$ . 
The values of the parameters $ \lambda, a$ and  $b$ which give the required curvature perturbation and spectral index are shown in the Table (\ref{values3_J}).
The slow roll parameters are
found to be $\epsilon_1 \simeq - \epsilon_3 \simeq 0.007$ and $\epsilon_2 \simeq - \epsilon_4 \simeq -0.013$ for chosen range of $\lambda$.  The calculated 
value for the  tensor to scalar ratio and Hubble parameter  are $r\simeq0.002 $ and $H \sim 10^{-3}  M_p$ respectively.

 The values of $F=1+\frac{\xi b \Phi^a R^{b-1}}{M_p^{a+2b-2}}$  are found to be $\sim  10^5$ i.e  much larger than unity and hence our assumption of 
dropping the unity in the expression for $F$ is justified.
Also we find that the order of the term $\frac{\kappa^2 \dot{\phi}^2}{F H^2} \sim 10^{-9} $ is much smaller than $6 \epsilon_3^2\sim 10^{-4}$
as assumed in  section (\ref{scal_pert}).

 We find that  in the limit $a\simeq2$ and $b\simeq1$ the correct value of $\Delta_{\mathcal R}^2$ and $n_{\mathcal R}$ are
obtained for $\lambda \sim 0.1$ only for large value of $\xi \sim 10^4$. Our Jordan frame calculation in this limit is
consistent with the results of \cite{Bezrukov,Bezrukov1,Bezrukov2,DeSimone:2008ei} who do the calculation in the Einstein frame.

\begin{table}
\center
    \begin{tabular}{|l|l|l|l|l|l|}
        \hline
       $ \ \lambda$ & $ \ 0.1  $  &  $ \ 10^{-2}  $  &   $ \ 10^{-3} $ & $ \ 10^{-4} $ & $\ 10^{-5}$ \\ \hline
        \ $\phi_{f} |_{_{(\phi_E=1 M_p)}}$  &$ \ 0.0146 M_p $ & $\ 0.0253 M_p$  & $ \ 0.044 M_p $ & $\ 0.077 M_p$&$\ 0.134 M_p$\\ \hline
         \ $\phi_{J}|_{_{(\phi_E=13 M_p)}}$ &$ \ 3.566 M_p $ & $\ 6.187 M_p $  & $ \ 10.77 M_p $ & $\ 18.8 M_p$&$\ 32.77 M_p$\\ \hline
       \ a &$ \ 3.56398962 $ & $\ 3.27512990$  & $ \ 3.02576940 $ & $\ 2.80956100$&$\ 2.62085100$\\ \hline
       \ b & $ \ 0.21800513 $ & $\ 0.36243484$  & $\ 0.48711456$& $\ 0.59521700$&$\ 0.68956620$ \\ \hline
       \ a+2b & $ \ 3.999999 $ & $ \ 3.999999$ & $ \ 3.999998 $ & $ \ 3.999995 $ & $ \ 3.99998 $ \\ \hline 
    \end{tabular}
    \caption{The values of parameters (a,b) are evaluated in the Jordan frame 
    at  $\xi=1$ and $\phi_J|_{_{\phi_E=13 M_p}}$ for different values of $\lambda$.}
    \label{values3_J}
\end{table}

\section{Discussion and Conclusion} \label{concl}

We have generalised the curvature coupling models of Higgs inflation to study inflation with  a scalar field for a
$\frac{\lambda}{4} \Phi^4$ potential and a curvature coupling of the form $ \frac{\xi \Phi^a R^b}{M_p^{a+2b-4}} $. It may be possible to generate a tree level term of this form
by choosing a suitable Kahler potential in a $f(\cal R)$ supergravity theory \cite{Ketov1,Ketov2,Ketov3}. 

We find that for $\xi =1$ and $\lambda$ in the range $(10^{-5}-0.1)$, the phenomenologically
acceptable parameters $a$ and $b$ fall in the ranges $(2.3 -3.6) $ and $(0.77-0.22)$ respectively. We discover an interesting symmetry related to the numerical value of $a$ and $b$
which give the correct amplitude and spectral index. We find that for any value of $\lambda$ the values of $a$ and $b$ which give the required density perturbations
satisfy the relation $(a+2b) \simeq 4$ as shown in Table(\ref{values3_J}). This means that the curvature coupling term $ \frac{\xi \Phi^a R^b}{M_p^{a+2b-4}} $ has no dimensional couplings and
is scale invariant.


It has been shown that the Higgs self coupling can go from $\lambda = 0.13$ at the electroweak scale for
the 125 GeV Higgs to $\lambda =10^{-5}$ at the Planck scale by tuning the top mass or by introducing extra
interactions \cite{Holthausen,Degrassi,Chakrabortty:2012np}. This leads us to conclude that the Higgs field
may still be a good candidate for being the inflaton in the early universe if one considers a generalised curvature-Higgs
coupling of the form $\xi \Phi^a R^b$.

The tensor to scalar ratio $r$ in this model is small and
the $\frac{\lambda}{4} \Phi^4$ with scalar curvature couplings is not ruled out by observational limits on $r$ unlike
the pure  $\frac{\lambda}{4} \Phi^4$ theory \cite{WMAP9,Boubekeur:2012xn}.

  We find that the values of $(a, b) $ computed with Jordan and Einstein frame calculations of the curvature perturbation and spectral
index are comparable but are not identical because unlike the $\xi \Phi^2 R$ theory, in the $\xi \Phi^a R^b$ theory it is not possible 
in general to go to an Einstein frame with a conformal transformation. If the space is quasi de-Sitter however such an transformation 
given by Eq. (\ref{Omega_eqn1}) is possible
but the results will differ in the two frames due to the slow roll approximation. Finally, by requiring the curvature coupling parameter to be of order unity, we have
evaded the problem of unitarity violation in scalar-graviton scatterings \cite{Hertzberg:2010dc} which
plagued the $\xi \Phi^2 R$ Higgs inflation models  \cite{Bezrukov,Bezrukov1,Bezrukov2,DeSimone:2008ei}.

\begin{center}
 {\bf Acknowledgment}
 \end{center}
We thank the anonymous referees for their constructive suggestions.


\begin{thebibliography}{unsrt}
\bibitem{starobinsky1} A.~A.~Starobinsky,
  JETP Lett.\  {\bf 30}, 682 (1979)
  [Pisma Zh.\ Eksp.\ Teor.\ Fiz.\  {\bf 30}, 719 (1979)].
  
\bibitem{starobinsky2} A.~A.~Starobinsky, Phys.\ Lett.\ B {\bf 91}, 99 (1980).


\bibitem{kazanas}   D.~Kazanas,
  Astrophys.\ J.\  {\bf 241}, L59 (1980).


\bibitem{sato}
K.~Sato,
  Mon.\ Not.\ Roy.\ Astron.\ Soc.\  {\bf 195}, 467 (1981).


\bibitem{guth}
  A.~H.~Guth,
  Phys.\ Rev.\ D {\bf 23}, 347 (1981).


\bibitem{linde1}
 A.~D.~Linde,
  Phys.\ Lett.\ B {\bf 108}, 389 (1982).
 
\bibitem{linde2}
 A.~D.~Linde,
 Phys.\ Lett.\ B {\bf 114}, 431 (1982).

 \bibitem{linde3}
 A.~D.~Linde, Phys.\ Lett.\ B {\bf 116}, 335 (1982).
\bibitem{albrecht}
 A.~Albrecht and P.~J.~Steinhardt,
  Phys.\ Rev.\ Lett.\  {\bf 48}, 1220 (1982).

\bibitem{Lyth:1998xn}
  D.~H.~Lyth and A.~Riotto,
    Phys.\ Rept.\  {\bf 314}, 1 (1999)  [hep-ph/9807278].  

\bibitem{WMAP9}
  C.~L.~Bennett, D.~Larson, J.~L.~Weiland, N.~Jarosik, G.~Hinshaw, N.~Odegard, K.~M.~Smith and R.~S.~Hill {\it et al.},
   arXiv:1212.5225 [astro-ph.CO].  


\bibitem{cms}
  S.~Chatrchyan {\it et al.}  [CMS Collaboration],
   Phys.\ Lett.\ B {\bf 716}, 30 (2012)  [arXiv:1207.7235 [hep-ex]].  


\bibitem{Atlas}
  G.~Aad {\it et al.}  [ATLAS Collaboration],
   Phys.\ Lett.\ B {\bf 716}, 1 (2012)  [arXiv:1207.7214 [hep-ex]].  

\bibitem{Sher}
  M.~Sher,
   Phys.\ Lett.\ B {\bf 317}, 159 (1993)  [Addendum-ibid.\ B {\bf 331}, 448 (1994)]  [hep-ph/9307342].  

\bibitem{Froggatt}
  C.~D.~Froggatt and H.~B.~Nielsen,
   Phys.\ Lett.\ B {\bf 368}, 96 (1996)  [hep-ph/9511371].  

\bibitem{Espinosa}
  J.~R.~Espinosa and M.~Quiros,
   Phys.\ Lett.\ B {\bf 353}, 257 (1995)  [hep-ph/9504241].
\bibitem{Schrempp}
  B.~Schrempp and M.~Wimmer,
   Prog.\ Part.\ Nucl.\ Phys.\  {\bf 37}, 1 (1996)  [hep-ph/9606386].

\bibitem{Holthausen}
  M.~Holthausen, K.~S.~Lim and M.~Lindner,
  JHEP {\bf 1202}, 037 (2012)  [arXiv:1112.2415 [hep-ph]].  
\bibitem{Degrassi}
  G.~Degrassi, S.~Di Vita, J.~Elias-Miro, J.~R.~Espinosa, G.~F.~Giudice, G.~Isidori and A.~Strumia,
    JHEP {\bf 1208}, 098 (2012)  [arXiv:1205.6497 [hep-ph]].


\bibitem{Isidori:2007vm}
  G.~Isidori, V.~S.~Rychkov, A.~Strumia and N.~Tetradis,
    Phys.\ Rev.\ D {\bf 77}, 025034 (2008)  [arXiv:0712.0242 [hep-ph]].  

\bibitem{Masina:2011un}
  I.~Masina and A.~Notari,
  Phys.\ Rev.\ Lett.\  {\bf 108}, 191302 (2012)  [arXiv:1112.5430 [hep-ph]].

\bibitem{Masina:2011aa}
  I.~Masina and A.~Notari,
  Phys.\ Rev.\ D {\bf 85}, 123506 (2012)  [arXiv:1112.2659 [hep-ph]].  


\bibitem{Masina}
  I.~Masina,
   arXiv:1209.0393 [hep-ph].  
   
\bibitem{Spokoiny:1984bd} 
  B.~L.~Spokoiny,
  Phys.\ Lett.\ B {\bf 147}, 39 (1984).  
  
\bibitem{Futamase:1987ua} 
  T.~Futamase and K.~-i.~Maeda,
  Phys.\ Rev.\ D {\bf 39}, 399 (1989).  

\bibitem{Salopek:1988qh} 
  D.~S.~Salopek, J.~R.~Bond and J.~M.~Bardeen,
  Phys.\ Rev.\ D {\bf 40}, 1753 (1989).  

 \bibitem{Fakir:1990eg}
  R.~Fakir and W.~G.~Unruh,
    Phys.\ Rev.\ D {\bf 41}, 1783 (1990).  

\bibitem{Kaiser:1994vs}
  D.~I.~Kaiser,
  Phys.\ Rev.\ D {\bf 52}, 4295 (1995)  [astro-ph/9408044].  


\bibitem{Komatsu:1997hv}
  E.~Komatsu and T.~Futamase,
   Phys.\ Rev.\ D {\bf 58}, 023004 (1998)  [astro-ph/9711340].  


\bibitem{Komatsu:1999mt}
  E.~Komatsu and T.~Futamase,
  Phys.\ Rev.\ D {\bf 59}, 064029 (1999)  [astro-ph/9901127].  



\bibitem{Bezrukov}
  F.~L.~Bezrukov and M.~Shaposhnikov,
   Phys.\ Lett.\ B {\bf 659}, 703 (2008)  [arXiv:0710.3755 [hep-th]].  

   \bibitem{Barvinsky:2008ia} 
  A.~O.~Barvinsky, A.~Y.~Kamenshchik and A.~A.~Starobinsky,
  JCAP {\bf 0811}, 021 (2008)
  [arXiv:0809.2104 [hep-ph]].  
    
   
\bibitem{Bezrukov1}   F.~L.~Bezrukov, A.~Magnin and M.~Shaposhnikov,
  Phys.\ Lett.\ B {\bf 675}, 88 (2009)
  [arXiv:0812.4950 [hep-ph]].
  

 \bibitem{Bezrukov2}
  F.~Bezrukov, A.~Magnin, M.~Shaposhnikov and S.~Sibiryakov,
  JHEP {\bf 1101}, 016 (2011)
  [arXiv:1008.5157 [hep-ph]].
  
  
   \bibitem{DeSimone:2008ei}
  A.~De Simone, M.~P.~Hertzberg and F.~Wilczek,
  Phys.\ Lett.\ B {\bf 678}, 1 (2009)
  [arXiv:0812.4946 [hep-ph]].

\bibitem{Barvinsky:2009fy} 
  A.~O.~Barvinsky, A.~Y.~Kamenshchik, C.~Kiefer, A.~A.~Starobinsky and C.~Steinwachs,
  JCAP {\bf 0912}, 003 (2009)
  [arXiv:0904.1698 [hep-ph]].

  
\bibitem{Hertzberg:2010dc}
  M.~P.~Hertzberg,
  JHEP {\bf 1011}, 023 (2010)
  [arXiv:1002.2995 [hep-ph]].


 \bibitem{Germani:2010gm}
  C.~Germani and A.~Kehagias,
  Phys.\ Rev.\ Lett.\  {\bf 105}, 011302 (2010)
  [arXiv:1003.2635 [hep-ph]].

\bibitem{Barvinsky:2009ii} 
  A.~O.~Barvinsky, A.~Y.~Kamenshchik, C.~Kiefer, A.~A.~Starobinsky and C.~F.~Steinwachs,
  Eur.\ Phys.\ J.\ C {\bf 72}, 2219 (2012)
  [arXiv:0910.1041 [hep-ph]].

  
  \bibitem{giudice}  G.~F.~Giudice and H.~M.~Lee,
  Phys.\ Lett.\ B {\bf 694}, 294 (2011)
  [arXiv:1010.1417 [hep-ph]].

\bibitem{lerner1}
  R.~N.~Lerner and J.~McDonald,
  Phys.\ Rev.\ D {\bf 82}, 103525 (2010)
  [arXiv:1005.2978 [hep-ph]].


\bibitem{bauer}
 F.~Bauer and D.~A.~Demir,
  Phys.\ Lett.\ B {\bf 698}, 425 (2011)
  [arXiv:1012.2900 [hep-ph]].


\bibitem{Qiu}
  T.~Qiu and D.~Maity,
   arXiv:1104.4386 [hep-th].  

 
\bibitem{Atkins:2010yg} 
  M.~Atkins and X.~Calmet,
  Phys.\ Lett.\ B {\bf 697}, 37 (2011)
  [arXiv:1011.4179 [hep-ph]].
  
\bibitem{Mooij:2011fi}
  S.~Mooij and M.~Postma,
   JCAP {\bf 1109}, 006 (2011)  [arXiv:1104.4897 [hep-ph]].  

\bibitem{Kaiser:2010yu}
  D.~I.~Kaiser and A.~T.~Todhunter,
  Phys.\ Rev.\ D {\bf 81}, 124037 (2010)
  [arXiv:1004.3805 [astro-ph.CO]].

\bibitem{Greenwood:2012aj}
  R.~N.~Greenwood, D.~I.~Kaiser and E.~I.~Sfakianakis,
   arXiv:1210.8190 [hep-ph].  

\bibitem{shinji}
  A.~De Felice and S.~Tsujikawa,
  Living Rev.\ Rel.\  {\bf 13}, 3 (2010)  [arXiv:1002.4928 [gr-qc]].  


\bibitem{Chakrabortty:2012np}
  J.~Chakrabortty, M.~Das and S.~Mohanty,
  arXiv:1207.2027 [hep-ph].
  
\bibitem{Planck_group} Planck Collaboration, [arXiv:1303.5076].

\bibitem{Gong:2011qe}
  J.~-O.~Gong, J.~C.~Hwang, W.~-I.~Park, M.~Sasaki and Y.~-S.~Song,
   JCAP {\bf 1109}, 023 (2011)  [arXiv:1107.1840 [gr-qc]].  

\bibitem{Basak:2012pc}
  A.~Basak and J.~R.~Bhatt,
    arXiv:1208.3298 [hep-ph].  


\bibitem{prokopec}

J.~Weenink and T.~Prokopec,
  Phys.\ Rev.\ D {\bf 82}, 123510 (2010)
  [arXiv:1007.2133 [hep-th]].


\bibitem{hwang}
  J.~C.~Hwang and H.~Noh,
  Phys.\ Rev.\ D {\bf 65}, 023512 (2001)
  [astro-ph/0102005].



\bibitem{hwang1}
 J.~C.~Hwang and H.~Noh,
  Phys.\ Rev.\ D {\bf 71}, 063536 (2005)
  [gr-qc/0412126].



\bibitem{Boubekeur:2012xn}
  L.~Boubekeur,
  arXiv:1208.0210 [astro-ph.CO].

  \bibitem{Ketov1} 
  S.~J.~Gates, Jr. and S.~V.~Ketov,
  Phys.\ Lett.\ B {\bf 674}, 59 (2009)
  [arXiv:0901.2467 [hep-th]].
  
   \bibitem{Ketov2} 
  S.~V.~Ketov,
  arXiv:1309.0293 [hep-th].
  
  \bibitem{Ketov3} 
  S.~V.~Ketov,
  Phys.\ Lett.\ B {\bf 692}, 272 (2010)
  [arXiv:1005.3630 [hep-th]].
  

  
 
  


\end{thebibliography}
\end{document}